# Infrared photo-response of Fe-shunted Ba-122 thin film microstructures

L. Rehm, D. Henrich, M. Hofherr, S. Wuensch, *Member, IEEE*, P. Thoma, A. Scheuring, K. Il'in, M. Siegel, S. Haindl, K. Iida, F Kurth, B. Holzapfel, and L. Schultz

*Abstract*—We present a study of the response to pulsed infrared radiation of Fe-layer shunted pnictide thin film microstructures. The thin film multilayer consisting of 20 nm thick Fe-buffer, 50 nm thick Ba(Fe,Co)$_2$As$_2$ film and gold protection layer were deposited on heated MgO and MgAl$_2$O$_4$ substrates by pulsed-laser deposition. The multilayers were patterned into 5 to 8 µm wide and 5 µm long microbridges by electron-beam lithography and ion-milling technique. The microbridges show $T_c \approx 20$ K and a critical current density up to 2.56 MA/cm$^2$ at $T = 10$ K. The photo-response of Fe-shunted Ba(Fe,Co)$_2$As$_2$ thin film microbridges to infrared radiation was studied in a wide range of incident optical power, operation temperature and bias current. We have found that the electron energy relaxation in studied multilayers is dependent on substrate material and is 1.75 times faster in case of MgAl$_2$O$_4$ characterized by lattice matching to pnictide film in comparison to MgO substrate.

*Index Terms*—barium compounds, high temperature superconductors, infrared detectors, superconducting photodetectors.

## I. Introduction

IN 2008 a new class of superconducting materials based on iron (Fe) and arsenide (As), also referred to as pnictides, has been discovered [1]. One of these iron-based superconductor families which was successfully grown as thin films with thicknesses below $d = 100$ nm is the $AE$Fe$_2$As$_2$ (where $AE$ stands for an alkali earth element). This family, which was named after the structural formula (122), shows solid temporal and mechanical stability as well as high critical temperature $T_c$ (up to 38 K in a hole-doped bulk sample [2], 22 K for electron-doping [3]) and critical current densities $j_c$ (up to 4 MA/cm$^2$ for a 250 nm [4] and 3 MA/cm$^2$ for a 90 nm thick pnictide film [5]). The introduction of an additional iron (Fe) buffer between substrate and the superconducting film has been shown to further improve the Ba-122 texture quality and thus its the superconducting properties [6,7]. This allows further decrease of film thickness that is important for development of devices for different cryogenic applications like fast and sensitive bolometer detectors for electromagnetic radiation. Moreover, ferromagnetic/superconductor bi-layers of conventional and high-$T_c$ superconductors demonstrate a faster photo response in respect to detectors made from bare superconducting films [8].

In spite of several publications on studies of photo-response of pnictide bulk samples [9,10], there is lack of results obtained on micrometer and sub-micrometer sized structures made from pnictide thin films. Nevertheless, a decrease in lateral dimensions is required for improvement of the bolometer sensitivity, which is proportional to the detector volume [11] and for embedding of detectors into planar antenna structures for applications in THz spectra range [12].

In this paper, we present experimental investigations of the photo-response of patterned Fe/Ba-122 bi-layers deposited on different types of substrate. In section II, the deposition of the Fe/Ba-122 multilayer and the patterning of µm-sized samples is presented. Results on study of the pnictide samples response to infrared radiation with respect to bias current, operation temperature, geometry and substrate material are presented and discussed in section III.

## II. Technology

We investigate two sample architectures with an iron (Fe) buffered substrate, a Ba-122 pnictide layer and a gold (Au) protection layer. The samples vary in their substrates and different layer thicknesses of the Au. The first sample hereinafter referred to as sample A, was deposited on magnesium oxide (MgO) substrate and was covered with $d_{Au} \approx 100$ nm thick Au layer. The second sample hereinafter referred to as sample B, was grown on spinel (MgAl$_2$O$_4$) substrate and covered with a thinner ($\approx 20$ nm) Au layer.

### A. Ba(Fe$_{0.92}$Co$_{0.08}$)As$_2$ Thin Film Deposition

Prior to the thin film deposition, the substrates (MgO (100) and MgAl$_2$O$_4$ (100)) were heated to 1000 °C for the purpose of cleaning. The following epitaxial Fe buffer layer ($d_{Fe} = 20$ nm) was prepared by a two-step process: First, the Fe layer was deposited on the substrate by pulsed-laser deposition (PLD) with a laser repetition rate of 5 Hz at room temperature using a pure metal Fe target. Second, the sample was annealed at 750 °C.

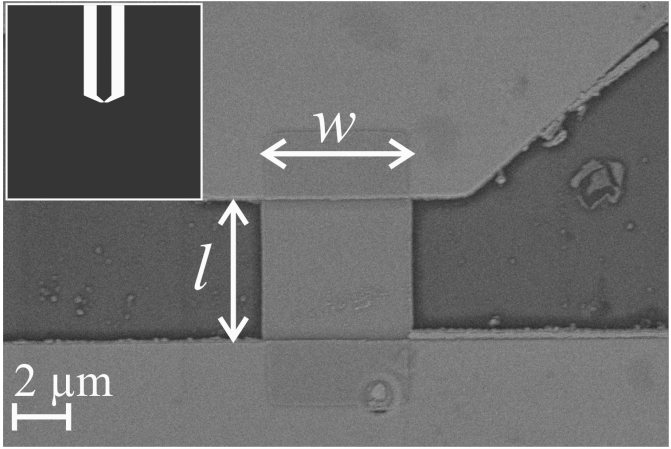

Fig. 1. SEM image of the patterned microstructure on for the film grown on MgO and $w = 5$ μm of the detection area. The dark areas are the substrate surface, the top and bottom light areas are the surface of the hardmask. The markers show the width $w$ and the length $l$ of the area free of Au, where the bare Ba-122 surface can be seen. Inset: Schematic sketch of the coplanar design.

The Ba(Fe$_{0.92}$,Co$_{0.08}$)$_2$As$_2$ thin films ($d_{\text{Ba-122}} \approx 50$ nm) of sample A and B were grown on Fe-buffered substrates by PLD. A sintered Co-doped Ba-122 target was ablated with a laser repetition rate of 7 Hz and at a deposition temperature of 750 °C. Finally, an Au protection layer to cover the Ba-122 was deposited also by PLD at room temperature.

All processes were conducted *in-situ* under UHV condition. A KrF excimer laser (wavelength 248 nm) with an energy density of 3-5 J/cm$^2$ was employed.

*B. Patterning Process*

It has been shown already that high temperatures may cause out-diffusion of As thus resulting in degradation of pnictide films. Therefore the process temperatures have to be below 100 °C and the chemical processing times have to be as short as possible to prevent also the degradation of the Ba-122 film. To pattern the films into μm-sized detector structures with a coplanar design (Fig. 1 inset), we developed a multi-step process with negative electron-beam lithography and ion-milling etching technique. The coplanar design is necessary to read out the high frequency detector response.

After surface cleaning with acetone and isopropanol, the film was coated with a negative electron-beam resist layer (≈ 380 nm), which allowed pre-baking at relatively low temperature of 85 °C. Afterwards the coplanar design of the detector structure and the length $l = 5$ μm of the detector area were defined by electron-beam lithography. On top of the sample a tri-layer was co-deposited. This tri-layer acts as a hardmask for the following etching process to transfer the coplanar design into the Ba-122. The multilayer consists of a thin adherent layer of $d \approx 10$ nm thick Nb, an Au layer ($d \approx 80$ nm) and on top a 70 nm thick NbN layer. The NbN layer is functioning as a sacrificial layer and the Au layer enables the contact improvement of the device and could also be used to implement a planar antenna for further THz investigations with thin film pnictide detectors [12].

After lift-off, the sample was coated a second time with the same electron-beam resist. The following electron-beam lithography defined the width $w$ of the detection area and varies between 5 and 8 μm. The developed resist together with the NbN layer mentioned above function as mask for etching, in which argon ions accelerated by an applied voltage of $U = 200$ V remove the unprotected film regions. After etching, the remaining resist was removed and the Au protection layer at the detection area was etched away with a second ion milling process (applied acceleration voltage $U = 120$ V). Here, the etching time has to be carefully controlled to avoid over-etching and damaging the underneath lying Ba-122 layer. The final structure is shown in Fig. 1. Both samples A and B were processed the same way. The fabricated device based on sample A (hereinafter referred to as A5) was fabricated with a detection area width of 5 μm. Based on sample B, two devices were patterned with different widths of 6 and 8 μm (hereinafter referred to as B6 and B8).

*C. Superconducting Properties*

Right after fabrication the sample were wire bonded in a standard four-probe configuration to measure the temperature dependence of the resistance $R(T)$ in the temperature range from 4.2 K to room temperature (inset in Fig. 2). During the measurements, the microstructures were biased with a constant current below 1 μA, which was small enough to avoid overheating of samples. The resistance of samples decreases with temperature shows ratio $R(295 \text{ K})/R(30 \text{ K}) = 1.8$ for sample A and 1.6 for sample B, comparable to what has been reported for microbridges with similar width made from 90 nm Ba-122 on LSAT substrate (≈ 1.9) [5].

Fig. 2 illustrates the normalized $R(T)$ in detail for a temperature range from 12 to 30 K. The width of transition $\Delta T$ was defined as the temperature difference between 10 % and 90 % of the normal state resistance $R_N$, taken at 30 K. The temperature at which the resistance decreases below 1 % of $R_N$ was defined as a critical temperature $T_c$. The values for $\Delta T$ and $T_c$ for all presented samples are shown in Table I. Although the $T_c$ value of the deposited pnictide films on MgAl$_2$O$_4$ was higher with respect to MgO sample, $T_c$ of the fabricated devices based on MgAl$_2$O$_4$ decreased significantly. The reduction of $T_c$ with decreasing width of microbridges can be attributed to the Ba-122 film damage at the sample edges [13]. Furthermore, the transition widths $\Delta T$ of B6 and B8 are strikingly increased compared to A5 and a two-step transition becomes apparent, indicating suppressed superconductivity of the bare detector area. This degradation is most likely due to micro-cracks in the ≈ 20 nm thick Au protection layer that were observed before the etching step. As a result, the micro-cracks enabled the partial bombardment of the Ba-122 with Ar ions during the second ion milling process to remove the Au layer. Therefore the uniform and dense Au layer is necessary to effectively protect the Ba-122 film during multi-step fabrication process.

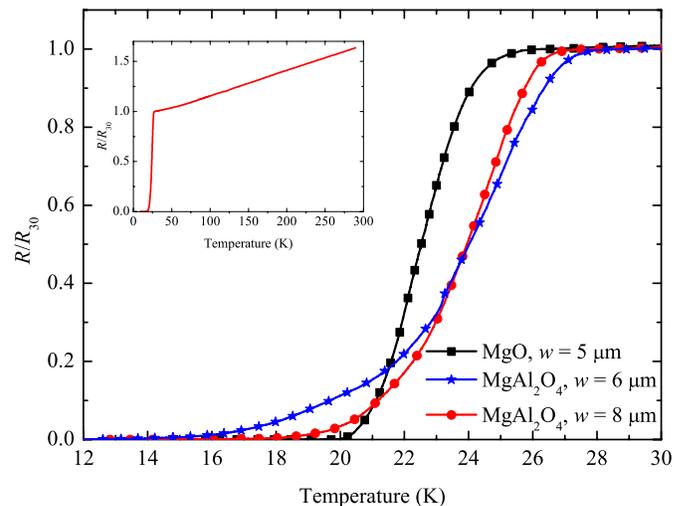

Fig. 2. Superconducting transition curves of the microstructures on different substrates: MgO—squares and MgAl$_2$O$_4$—stars, circles. The correspondent widths ($w$) of the detection area are indicated in the graph. The film on MgO has a 100 nm thick Au protection layer. The films on MgAl$_2$O$_4$ are covered only by a 20 nm thick Au layer. Inset: Superconducting transition curve of the film grown on MgAl$_2$O$_4$ and $w = 8$ μm from 4.2 K to room temperature.

TABLE I
PROPERTIES OF FE/BA-122 (20/50 NM) FILMS AND DETECTOR SAMPLES.

| substrate | $d_{Au}$ (nm) | $T_c$ (K) | $\Delta T$ (K) | sample | $w$ (μm) | $T_c$ (K) | $\Delta T$ (K) |
|---|---|---|---|---|---|---|---|
| | | before patterning | | | | after patterning | |
| MgO | 100 | 21.58 | 0.95 | A5 | 5 | 20.29 | 2.96 |
| MgAl$_2$O$_4$ | 20 | 27.28 | 1.07 | B6 | 6 | 15.76 | 6.69 |
| | | | | B8 | 8 | 18.8 | 4.57 |

The current-voltage characteristics of the devices were measured in the current-bias mode at 10 K. All samples showed a smooth and in a wide temperature range non-hysteretic transition between the superconducting and the resistive state, which is shown in Fig. 3. This particular behavior can be described by the influence of the iron buffer layer that short-circuits the Ba-122 in the normal state [13]. The critical current $I_c$ was determined as the current at which the voltage deviates from zero by 100 μV. This criterion is given by the resolution of the used measurement system. The respective critical current densities are $j_c$ (10 K) = 3.69 MA/cm$^2$ for A5, 0.18 MA/cm$^2$ for B6 and 0.39 MA/cm$^2$ for B8. The values found for the critical current densities for B6 and B8 are comparable to the previously reported value of 0.4 MA/cm² for single crystals [14]. Beyond that, the critical current densities for A5 at 10 K is significantly higher to what has been reported for microbridges with the same width range based on 100 nm thick Ba-122 and Fe buffered MgO (1.71 MA/cm$^2$ at 4.2 K) [13]. This increase of the critical current density of sample A5 can be mainly attributed to the improved critical temperature.

### III. PHOTO RESPONSE

The photo response of the fabricated samples was measured with a liquid helium Dewar insert measurement setup. The sample temperature was adjusted by the dip depth of the probe in the liquid helium bath and the contact gas pressure in the insert. The femtosecond pulsed infrared (IR) radiation from the laser source ($\lambda$ = 1550 nm) was passed into the insert by an optical multimode fiber. The 105 μm core diameter of the fiber and ≈ 0.7 mm distance between fiber end and sample ensure a large enough light spot for uniform illumination of the detector. The optical power incident on the detection area was determined with a photodiode at room temperature. During the measurement the sample was biased with a DC current from a low-noise current source. The response to the IR radiation was amplified by a two-stage low noise amplifier (20 dB gain), which was especially designed for cryogenic applications [15]. After amplification, the signal was feeded out to a spectrum analyzer by a rigid stainless steel coaxial cable and a vacuum feed-through. The two-stage low noise amplifier limits the system bandwidth to a frequency of 12 GHz.

#### A. Sensitivity

The photo-response was measured at temperatures below $T_c$ and wide range of bias currents $I_b$. For a given bias current, the output power of the detector showed linear dependence on the incident laser power over two orders of magnitude. All measurements were realized with power levels inside this dynamic range ($P$ ≈ 0.11 nW relative to 1 μm$^2$ detector area).

The voltage response in dependence on bias current showed a specific maximum at the saddle point $I_b$ of the $IV$ curve for each temperature. This maximum was defined as the optimal operation point and the typical value of the corresponding optimal operation current was around 10 mA for microstructures, depending on the width of detection area. The general behavior was similar to the photo-response of YBaCuO or NbN films [16]. At the optimal bias point, sensitivities up to 1.77 μV/pJ for B6 and 0.35 μV/pJ for B8 were achieved for $T$ = 8 K assuming 100% absorption of radiation in Ba-122 film, which is of course a worst-case estimation. To determine exact sensitivity values, the actual absorption of a Ba-122 thin film for $\lambda$ = 1550 nm is required.

The dependence of the output power of the detector on temperature for both B6 and B8 samples is shown in Fig. 4. The output signals of both samples demonstrate similar increase with decreasing temperature. The signal response of 6 µm wide detector (sample B6) at same temperatures is about 0.16 µW higher than the response of the microstructure with $w = 8$ µm (sample B8). Based on the results shown in Fig. 4, reduction of the detection area of the device leads to an increased signal response, as would be expected from bolometer theory where the sensitivity $S$ is inversely proportional to the sample volume $V$.

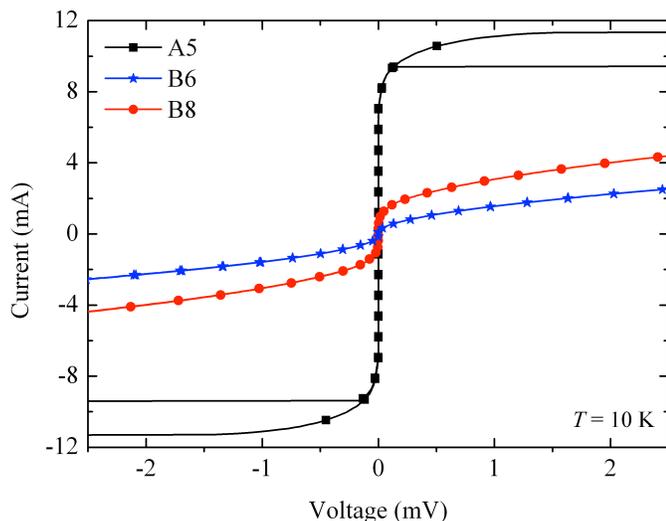

Fig. 3. Current-voltage characteristics of samples based on MgAl$_2$O$_4$ substrate: Stars—width of the detection area $w$ = 6 μm and circles— width of the detection area $w$ = 8 μm and on MgO substrate: Squares—width of the detection area $w$ = 5 μm measured at 10 K.

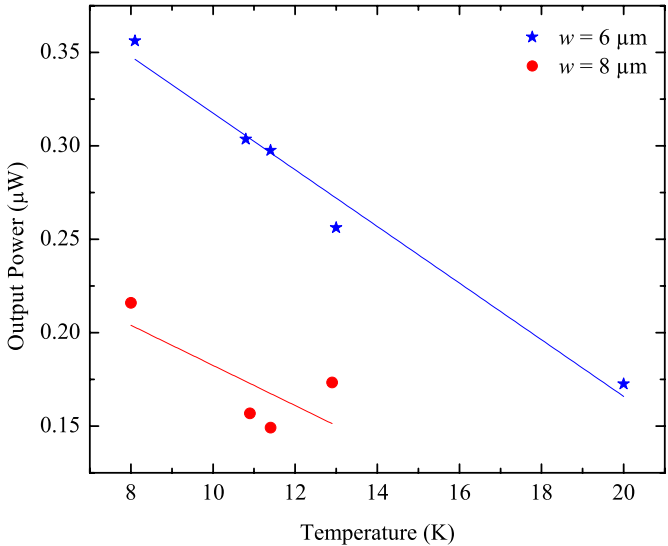

Fig. 4. Temperature dependence of the output power for different widths of the detection area: 6 µm—stars and 8 µm—circles. The radiation power on the detection area was ≈ 3 nW at λ = 1550 nm. The solid lines are to guide the eye.

### B. Energy Relaxation Time

The temporal evolution of a response pulse to the short IR laser excitation was studied using a 20 GHz sampling oscilloscope. Fig. 5 shows the responses of two pnictide microstructures on different substrates to pulsed IR radiation. The transients of the sample on MgO (Fig. 5(a)) and on $MgAl_2O_4$ (Fig. 5(b)) were acquired at the same temperature ($T \approx 10$ K). Both µm-sized pnictide detectors show similar behavior of the response transients. After a fast initial excitation ($\tau_{rise} \approx 540$ ps, which was limited by the readout electronics), the detectors show three-step voltage decay. The first two appearing during about 1.5 ns are with pretty steep unclear so far. After about two nanoseconds the films excited slopes similar to rising front of the pulse. Their origin is the by optical radiation [16] can be described in exponential relaxation tail is observed for both samples.

The energy relaxation processes in superconducting thin framework of the two-temperature model [17]. Here, it is assumed that the photon energy is first absorbed by a single quasi-particle and then distributed in the electronic sub-system of the material by an avalanche-like process. The slow part of the relaxation is governed by the energy transfer to the phonon subsystem and by the phonon escape to the substrate. In our case the situation becomes more complicated due to the Fe buffer layer between the superconducting film and the substrate. Thus, the escaping phonons face two interfaces with different transparencies. We identify the last part of the voltage pulse with the phonon escape from our Fe/Ba122 multilayers to substrate and fit it with an exponential decay to extract the relaxation time constant τ (solid lines in fig. 5). The corresponding values are τ = 1.75 ns for MgO and τ = 0.98 ns for $MgAl_2O_4$. This fits well to the observation of the about ≈ 5.7 K higher $T_c$ of as-deposited films on $MgAl_2O_4$ substrate (in comparison to MgO) which is caused by the improved lattice matching and in consequence leads to a more transparent phonon interface.

### IV. CONCLUSION

We have developed a reliable multi-step fabrication process for µm-sized detector structures based on a Fe/Ba-122/Au multi-layer system using electron beam lithography and ion milling. It has been found that the *in-situ* Au protection layer was crucial for the quality of final structures and needs to be uniform and dense to prevent the Ba-122 from degradation during the ion milling process.

Decrease of the detector volume of thin Fe/Ba122 films by decrease in width from 8 to 6 µm results in increase of the detector response while its temperature dependence was found to be independent of width.

The response of thin Fe/Ba122 film samples to pulsed infrared radiation shows multi-step relaxation and was studied using the time domain technique. The physical explanation of the fast relaxation steps is unclear so far. The slow bolometer tail of the response is most probably caused by the escape of non-equilibrium phonons from film to substrate. The characteristic time of this process for detectors made from films on $MgAl_2O_4$ substrate is almost two times shorter than for detectors with the same thickness of Fe/Ba122 multilayer deposited on MgO substrate. The difference in energy relaxation between samples on different substrates is caused by difference in lattice mismatch, which is larger for MgO.

### ACKNOWLEDGMENT

The work at IMS Karlsruhe is supported in part by the IEEE Council on Superconductivity. The work at IFW is supported by DFG under project HA 5934/3-1 (SPP 1458).

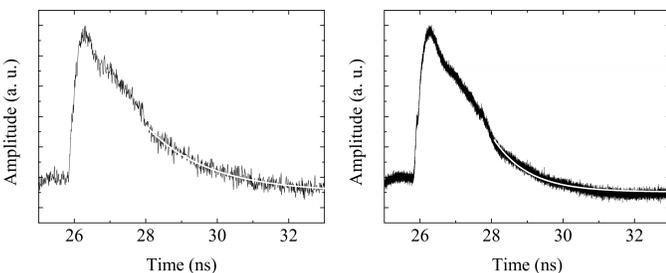

Fig. 5. Time-resolved voltage responses to pulsed radiation with λ = 1550 nm measured at 10 K. The samples differ in their substrates: left—MgO and right—$MgAl_2O_4$. The solid lines represent exponential fits to the tail part of the data.